\documentclass[12pt]{article}
\usepackage{amssymb}
\usepackage{amsmath}
\usepackage{latexsym}
\usepackage[dvips]{graphicx}
\def\ga{\mathrel{\raise.3ex\hbox{$>$\kern-.75em\lower1ex\hbox{$\sim$}}}}
\def\la{\mathrel{\raise.3ex\hbox{$<$\kern-.75em\lower1ex\hbox{$\sim$}}}}
\def\beq{\begin{equation}}
\def\eeq{\end{equation}}

\begin{document}
\begin{titlepage}
\pagestyle{empty}
\baselineskip=21pt
\rightline{\tt astro-ph/0603544}
\rightline{UMN--TH--2435/06}
\rightline{FTPI--MINN--06/07}
\rightline{March 2006}
\vskip 0.2in
\begin{center}
{\large{\bf Gravitational Waves from the First Stars}}
\begin{center}
\end{center}
\vskip 0.2in
{\bf Pearl Sandick}$^1$, {\bf Keith~A.~Olive}$^{2}$,  {\bf Fr\'ed\'eric~Daigne}$^3$,
and {\bf Elisabeth Vangioni}$^3$
{\it
$^1${Department of Physics, School of Physics and Astronomy,
University of Minnesota, Minneapolis, MN 55455 USA}\\
$^2${William I. Fine Theoretical Physics Institute, School of Physics and Astronomy,
University of Minnesota, Minneapolis, MN 55455 USA}\\
$^3${Institut d'Astrophysique de Paris, UMR 7095, CNRS, Universit\'e Pierre et Marie Curie-Paris VI, 98 bis bd Arago, F-75014, Paris, France}}

\vskip 0.1in

{\bf Abstract}
\end{center}
We consider the stochastic  background of gravitational waves 
produced by an early generation of Population III
stars coupled with a normal mode of star formation at lower redshift.
The computation is performed in the framework of 
hierarchical structure formation and is based
on cosmic star formation histories constrained to reproduce the observed star formation
rate at redshift $z \la 6$, the observed chemical abundances in damped Lyman alpha absorbers
and in the intergalactic medium, and to allow for an early reionization of the Universe
 at $z\sim 10-20$ as indicated by the first year results released by WMAP.
We find that the normal mode of star formation produces a gravitational wave background
which peaks at 300-500 Hz and is within LIGO III sensitivity.  The Population III
component peaks at lower frequencies (30-100 Hz depending on the model),
and could be detected by LIGO III as well as the planned BBO and DECIGO interferometers.

\baselineskip=18pt \noindent

\end{titlepage}

\section{Introduction}

In the last few years, there has been significant progress in our understanding of the 
early cosmic star formation history.
The first year data obtained by WMAP indicates a large optical depth, implying that the universe became reionized at high redshift in the range $ 11 < z < 30$ at 95\% CL~\cite{wmap}. To account for a period of early reionization, it has been argued that a generation of very massive stars preceded the oldest observed generation of Population II stars~\cite{reion}. In addition, cosmic star formation rate (SFR) at $z \lesssim 6$, has been observed at levels significantly larger than the current rate \cite{csfr}.  
Taken together, the evidence suggests that the distribution of the first stars (Population III) 
are described by a top-heavy initial mass function (IMF), formed in primordial metal-free structures with masses of order $10^7$ M$_{\odot}$. As these stars produced heavier elements, the universe achieved a critical metallicity ($\sim 10^{-4}$ times solar metallicity) \cite{zcrit}, at which point the massive mode of star formation yielded to a more normal distribution of stellar masses with a 
SFR peaked at $z \approx 3$. A more detailed understanding of the first epoch of star formation
will rely on the phenomenological consequences of the models such as element enrichment
and supernova rates \cite{others,dosva,dossv}.

One consequence of this new view of star formation is an enhanced rate of core 
collapse supernovae. The resultant relic neutrino background was 
investigated in Ref.~\cite{othersnu,dosv}. In each core collapse supernova explosion, 
the bulk of the energy released is in the form of neutrinos which, because 
they are weakly interacting, retain information about their origins. Although the 
neutrino background from a massive mode of star formation (Pop III) at early times is 
not likely to be detected due to its redshifted spectrum, the prospects for 
observation of the spectrum produced by the normal mode of star formation (Pop II)
in the near future are good. 

One possible probe of the massive mode, however, 
is the stochastic background of gravitational waves produced by cosmological core collapse supernovae, which we consider here. 
Although only a small fraction of the total energy of core collapse is emitted in gravitational radiation, the improved sensitivities by way of correlation of currently operating ground-based interferometers GEO600 \cite{geo}, LIGO II and III  \cite{ligo}, TAMA \cite{tama}, and VIRGO \cite{virgo}, and of future space-based antennas BBO \cite{BBO}, LISA \cite{lisa}, DECIGO \cite{seto}, 
make the positive detection of the accumulated gravitational wave background plausible.

The gravitational wave background from core collapse supernovae resulting in black holes has been calculated in Refs.~\cite{fms,araujoSH}, with estimates of the peak of the differential energy density spectrum of $\Omega_{GW} h^2 = 10^{-11} - \textrm{few} \times 10^{-9}$ reaching its maximum value at frequencies anywhere from a few hundred Hertz to a few thousand Hertz. The calculation has also been made specifically for Population III supernovae resulting in black holes in Ref.~\cite{araujo}. They estimate the spectrum to peak at $\Omega_{GW} h^2 \approx 10^{-8}$ at a frequency of $O(100)$ Hz. More recently, the spectra from both a normal mode of star formation, in which all stars collapse to form neutron stars, and a Population III mode, in which all stars collapse to form black holes, was calculated in Ref.~\cite{bsrjm}. They find both peaks to be located at roughly $\Omega_{GW} h^2 \approx \textrm{few} \times 10^{-12}$ in the most optimistic case, with the peak frequency dependent on the redshift range over which gravitational collapse occurs. As these spectra are highly model dependent, and given that previous estimates of the differential closure density span four orders of magnitude in the range where detection may soon be possible, it is now important to examine the sensitivity of a detectable signal to the star formation history.
 
Here, we incorporate fully developed chemical evolution models which trace the history of pre-galactic structures as well as the IGM and are based on a $\Lambda$CDM cosmology with a Press-Schechter model of hierarchical structure formation \cite{ps}. We adopt the chemical evolution models of Daigne {\it et al.}~\cite{dossv} and consider several bimodal star formation histories, each with a normal component of star formation as well as a massive component describing Population III stars. Given an IMF and a respective SFR, we calculate the expected gravitational wave background and compare this
result with detector sensitivities.

\section{Calculation of the Gravitational Wave Background}
\label{sec:calc}

Gravitational waves can be characterized by a frequency, $f$, and an amplitude, $h$, which is defined by the degree of quadrupole anisotropy and strength of the source. Using an amplitude determined from the simulation of a 15 M$_{\odot}$ star from Ref.~\cite{sim}, Buonanno {\it et al.}~\cite{bsrjm} made a generalization to larger stars by using the function's dependence on the anisotropy and neutrino luminosity during collapse. The shape of the gravitational wave spectrum can then be described by the dimensionless quantity
\beq
f |\tilde{h}(f)| = \frac{G_N}{\pi c^4 D}E_{\nu} \langle q \rangle \Big( 1 + \frac{f}{a} \Big)^3 e^{-f/b},
\label{fh}
\eeq
where $\tilde h$ is the Fourier transform of $h$,  $G_N$ is Newton's constant, $E_{\nu}$ is the total energy emitted in neutrinos, $\langle q \rangle$ is the average value of the anisotropy parameter, $q$, defined in Ref.~\cite{mj97}, and $D$ is the distance to a typical supernova. Although it is necessary to know the distance in order to quantify the amplitude, we will see that the accumulated energy density does not depend on this parameter.  Once the structure of the amplitude is imposed, the constants $a$ and $b$ determine the specific spectral shape. By roughly reproducing the source spectrum from simulation model s15r in Ref.~\cite{sim}, one 
obtains $a \approx 200$ Hz.~and $b \approx 300$ Hz. 
The sensitivity to these choices will be discussed in section~\ref{sec:results}.

A stochastic background of gravitational waves with energy density $\rho_{GW}$ and frequency $f$ is best described by the differential closure density parameter~\cite{maggiore},
\beq 
\Omega_{GW}(f) = \frac{1}{\rho_c}\frac{d\rho_{GW}}{d\mathrm{log}f},
\eeq
with $\rho_c = 3H_0^2/8 \pi G_N$ the critical density. Given a star formation history, consisting of a star formation rate per comoving volume (SFR), $\psi(t)$, and an initial mass function (IMF), $\phi(m)$, and the gravitational wave amplitude, the stochastic background of gravitational waves produced by cosmological core collapse supernovae is
\beq
\Omega_{GW}(f) = \frac{16 \pi^2 c^3 D^2}{15 G_N \rho_c} \int_0^{z_i}\frac{dz}{1+z} \Bigg| \frac{dt}{dz} \Bigg| \int_{M_{min}}^{M_{max}} dm \phi(m) \psi(t-\tau(m)) f'^3 \vert \tilde{h}(f',m) \vert^2,
\eeq
where $f'$ is the frequency at emission, related to the observed frequency, $f$, by $f'=f(1+z)$, $z_i$ is the initial redshift at which supernovae begin to occur, $M_{min}$ and $M_{max}$ are the minimum and maximum masses in each model for which supernovae occur, and $\tau(m)$ is the lifetime of a star of mass $m$~\cite{bsrjm}. Using Eq.~\ref{fh}, one obtains
\beq
\Omega_{GW}(f) = \frac{16 G_N}{15 c^5 \rho_c} \int_0^{z_i}dz \Bigg| \frac{dt}{dz} \Bigg| \int_{M_{min}}^{M_{max}} dm \phi(m) \psi(t-\tau(m)) \langle q \rangle^2 E_{\nu}^2 f' \Big( 1 + \frac{f'}{a} \Big)^6 e^{-2f'/b}.
\label{omega}
\eeq

The parameter $\langle q \rangle$ must be determined from simulations. As in Ref.~\cite{bsrjm}, we take $\langle q \rangle = 0.45\%$ for progenitors that collapse to neutron stars, as determined by simulation in \cite{sim}. 
The total energy emitted in gravitational waves can be described in terms of the amplitude as
\beq 
\begin{split}
E_{GW}& = \frac{16 \pi^2 c^3 D^2}{15 G_N} \int df f^2 |\tilde{h}(f)|^2 \\
      & = \frac{16 G_N}{15 c^5} \langle q \rangle^2 E_{\nu}^2 \int df \Big( 1 + \frac{f}{a} \Big)^6 e^{-2f/b}.
\label{energy}
\end{split} 
\eeq
The efficiency of gravitational wave production, $\epsilon$,  is defined in terms of the  remnant mass, $M_r$, by
\beq
E_{GW}=\epsilon M_{r} c^2.
\label{efficiency}
\eeq
Using $E_{\nu} = 3 \times 10^{53}$ ergs, we determine the efficiency to be $\epsilon  = 1.5 \times 10^{-7}$
when a neutron star remnant is produced.

When collapse proceeds to form a black hole, we assume, as in Ref.~\cite{bsrjm}, an efficiency of gravitational wave production found by Fryer {\it et al.}~of $\epsilon = 2 \times 10^{-5}$~\cite{fwh}.
One can then determine the quantity $\langle q \rangle E_{\nu}$ for a star of mass $m$ by assuming only a spectral shape (here, Eq.~\ref{fh}) and an efficiency of gravitational wave production. In this case it is necessary to specify the mass of the remnant, but not the neutrino luminosity. Note that the
efficiency of gravitational wave production is much greater when a black hole rather than a neutron star is produced as a remnant.

\begin{figure}[ht]
\centering
\includegraphics[width=0.75\textwidth]{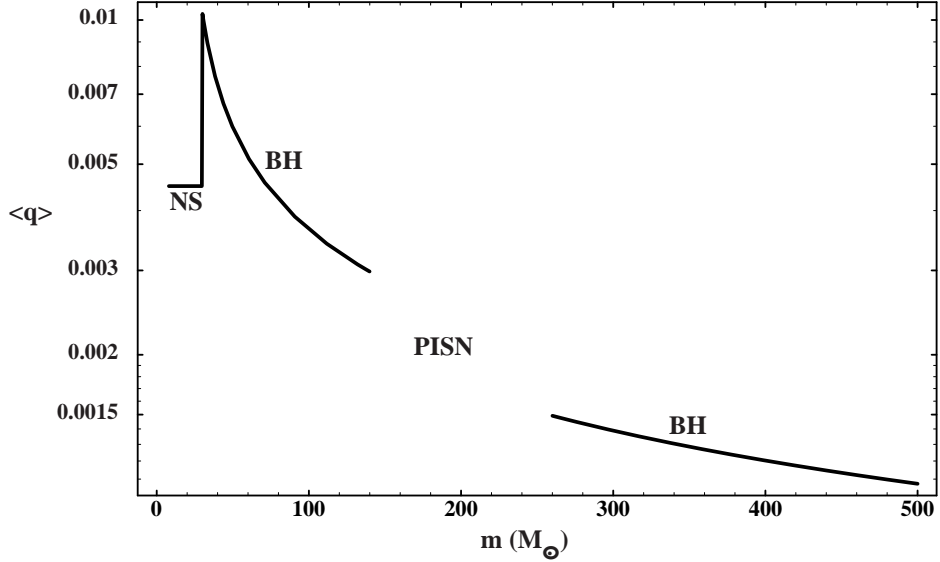}
\caption{Anisotropy parameter, $\langle q \rangle$, as a function of the mass of the progenitor. The energy emitted in neutrinos is taken to be $E_{\nu}=3 \times 10^{53}$ ergs for 8 M$_{\odot} < m <$ 30 M$_{\odot}$ and  $E_{\nu}=\frac{3}{10} M_r c^2$ for stars with  $m >$ 30 M$_{\odot}$ that collapse to black holes. For stars with 30 M$_{\odot} < m < 140$ M$_{\odot}$, the mass of the remnant is taken to be the mass of the helium core before collapse, $M_{He} = 13/24 \cdot (m-20 M_{\odot})$~\cite{hw02}. In the case of the larger stars with $m > 270$ M$_{\odot}$, $M_r=m$. Since $\langle q \rangle \propto 1/E_{\nu}$, this is a minimum average value for the anisotropy parameter for these larger stars.
\label{qplot}}
\end{figure}

One interesting consequence of choosing the efficiency of gravitational wave production to be constant for stars that collapse to black holes is that the anisotropy parameter actually decreases for larger stars, as can be seen in Fig.~\ref{qplot}. This is because the quantity $\langle q \rangle E_{\nu} \propto \sqrt{M_r}$, but the energy emitted in neutrinos itself, $E_{\nu} \propto M_r$, resulting in $\langle q \rangle \propto 1/\sqrt{M_r}$. The values obtained for $\langle q \rangle$ are not unreasonable for larger stars in Models 1 and 2b, however it has been noted that one might expect a larger degree of anisotropy due to rotation and/or violent explosions~\cite{bsrjm}. Very little is known about the anisotropy, however there is evidence that stars that collapse to black holes should be more efficient emitters of gravitational waves.  Stark and Piran found a maximum efficiency of $\epsilon \lesssim 7 \times 10^{-4}$ for an axisymmetric collapse resulting in a black hole~\cite{sp85}, and more recently, Fryer {\it et al.}~obtained the efficiency of $2 \times 10^{-5}$ for a 100 M$_{\odot}$ black hole remnant, which we adopt. Note that since $\Omega_{GW} \propto \epsilon$, if the efficiency is actually closer to Stark \& Piran's maximum value, the observed energy density in gravitational waves may be as much as 35 times larger than that shown in our results below, and even more if the collapse is less symmetric. While it is possible to fix the anisotropy parameter to be a function of mass, there is no data to support any particular functional dependence. Note that the gap in  Fig.~\ref{qplot} falls in the range where
pair-instability supernovae are important.  The gap reflects a lack of knowledge 
concerning the production of gravitational waves for these stars.

In all of our computations, we assume a flat $\Lambda$CDM cosmology with  
\beq
\left| \frac{dt}{dz} \right| = \frac{9.78\,h^{-1}\,\textrm{Gyr}}{(1+z)\sqrt{\Omega_{\Lambda}+\Omega_m(1+z)^3}}
\eeq
where $\Omega_{\Lambda}=0.73$, $\Omega_m=0.27$, and $h=0.71$~\cite{wmap}.

\section{Star Formation Models}
\label{sec:SF models}

The cosmic star formation histories considered here have been adopted from the detailed model of chemical evolution in Ref.~\cite{dossv}. The models are described by a bimodal 
birthrate function of the form
\beq
B(m,t,Z) = \phi_1(m)\psi_1(t) + \phi_2(m)\psi_2(Z)
\eeq
where $\phi_{1(2)}$ is the IMF of the normal (massive) component of star formation and $\psi_{1(2)}$ is the respective SFR. $Z$ is the metallicity. The normal mode contains stars with mass between 0.1 M$_{\odot}$ and 100 M$_{\odot}$ and has a SFR which peaks at  $z \approx 3$. The massive component dominates at high redshift. The IMF of both modes is taken to be a power law with a near Salpeter slope
so that,
\beq 
\phi_i (m) \propto m^{-(1+x) }
\label{IMF}
\eeq
 with $x=1.3$.
 Each IMF is normalized independently by
 \begin{equation}
\int_{m_\mathrm{inf}}^{m_\mathrm{sup}} dm\ m \phi_i(m)=1\ ,
\label{norm}
\end{equation}
differing only in the specific mass range of each model. 
Here we consider two different mass ranges for the massive mode, although three mass ranges are presented in Ref.~\cite{dossv}, as will be discussed presently.  Both the normal and massive components can contribute to the chemical enrichment of galaxy forming structures and the IGM, though the normal mode is not sufficient for accounting for the early reionization of the IGM \cite{dosva}.

Here, we restrict our attention to the best fit hierarchical model in \cite{dossv} in which the minimum 
mass for star formation is $10^7$ M$_\odot$. The normal mode SFR is given by
\beq
\psi_1(t) = \nu_{1} M_{struct} \exp{(-t/\tau_{1})}\ ,
\eeq
where  $\tau_{1} = 2.8$ Gyr is a characteristic timescale and $\nu_{1}= 0.2$ Gyr$^{-1}$ 
governs the efficiency of the star formation. In contrast, the massive mode SFR is defined by
\begin{equation}
\psi_2(t) = \nu_{2} M_{\rm ISM} \exp{\left(- Z_{\rm IGM} / Z_\mathrm{crit}\right)}\ ,
\end{equation}
with $\nu_{2}$ maximized to achieve early reionization without the overproduction of metals or 
the over-consumption of gas. We adopt $Z_\mathrm{crit}/Z_\mathrm{\odot}=10^{-4}$ .

We consider three different models, labeled Models 1, 2a, and 2b to describe the massive mode.
They are distinguished by their respective stellar mass ranges.
In Model 1,  the IMF is defined for stars with masses, 40 M$_{\odot} \le m \le 100$ M$_{\odot}$. 
All of these stars die in core collapse supernovae leaving a black hole remnant.
Model 2a is described by
very massive stars which become pair instability supernovae.  The IMF is defined for 140 M$_{\odot} \le m \le 260$ M$_{\odot}$. Finally, 
the most massive stars are considered in Model 2b and fall in the range 270 M$_{\odot} \le m \le 500$ M$_{\odot}$, with the SFR as in Model 1. These stars entirely collapse into black holes and do not contribute 
to the chemical enrichment of either the ISM or IGM.  The coefficient of star formation, $\nu_2$
is 80, 40, and 10 Gyr$^{-1}$ for Models 1, 2a, and 2b respectively.
In each case, star formation begins at very high redshift ($z \simeq 30$) but peaks 
at redshifts between 10 and 15, depending on the model.  Note that the absolute value of the SFR
depends not only on $\nu_2$, but also on the efficiency of outflow.  See \cite{dossv} for details.

\begin{figure}
\begin{center}
\resizebox{\textwidth}{!}{\includegraphics{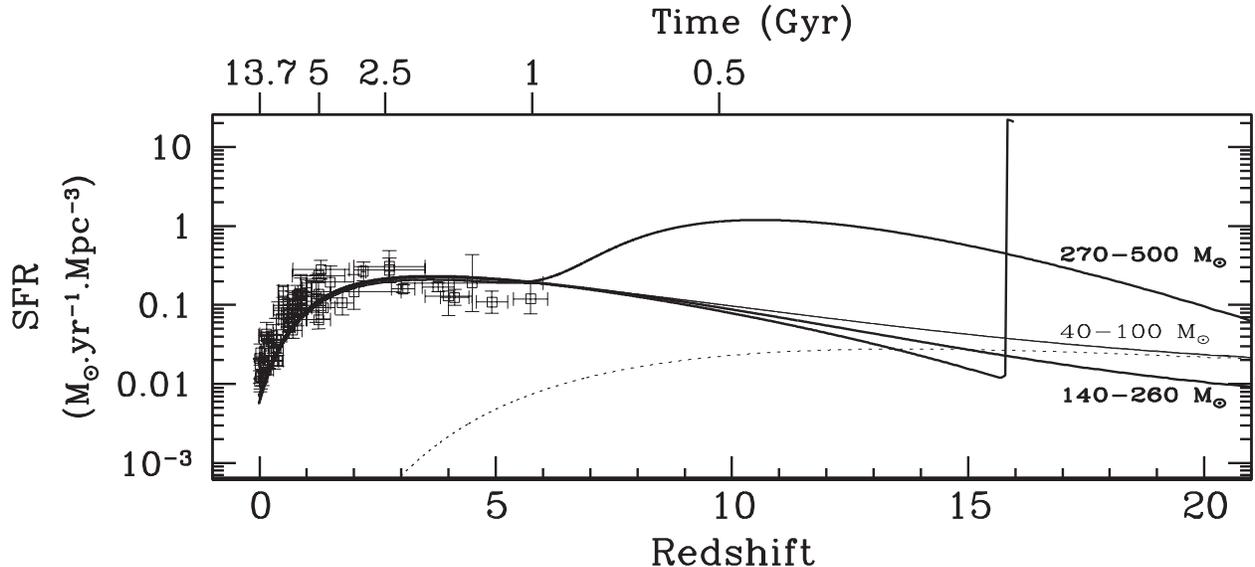}}
\end{center}
\caption{The star formation rate for Models 1, 2a, and 2b as labeled by their respective mass ranges, and a rapid burst model.
The dashed line in the upper panel shows the SFR of the massive mode of Model 1.}
\label{fig:extremModel1SFRFracBZ}
\end{figure}

In Figure~\ref{fig:extremModel1SFRFracBZ}, 
we show the SFR, for Models 1, 2a, and 2b (including the normal mode).
Also shown by the dotted curve is the SFR for the massive mode alone in Model 1.
For comparative purposes, we also consider an
example of a model (with an IMF as in Model 1) in which the massive mode
occurs as a rapid burst at $z = 16$ designated as Model 1e.
This is also shown in Figure~\ref{fig:extremModel1SFRFracBZ}
compared with the analogous result for Model 1.
As one can see, 
the massive burst SFR is significantly larger than the model considered above for the short duration of the burst, while at lower redshifts the SFR,
which is determined by $\nu_1$ is nearly identical.  
Because of our lack of understanding of gravitational wave production
in pair-instability supernovae, we will not consider Model 2a any further.

The rate of core collapse supernovae
can be calculated directly in terms of the IMF and SFR
\beq
SNR = \int_{max(8M_{\odot},m_{min(t)})}^{m_{sup}} dm \phi(m) \psi(t-\tau(m)),
\label{SNR}
\eeq
where $m_{min(t)}$ is the minimum mass of a star with lifetime less than $t$. The differential energy density parameter for each model is calculated using Eq.~\ref{omega}.

\section{Results}
\label{sec:results}

\begin{figure}[ht]
\centering
\includegraphics[width=0.75\textwidth]{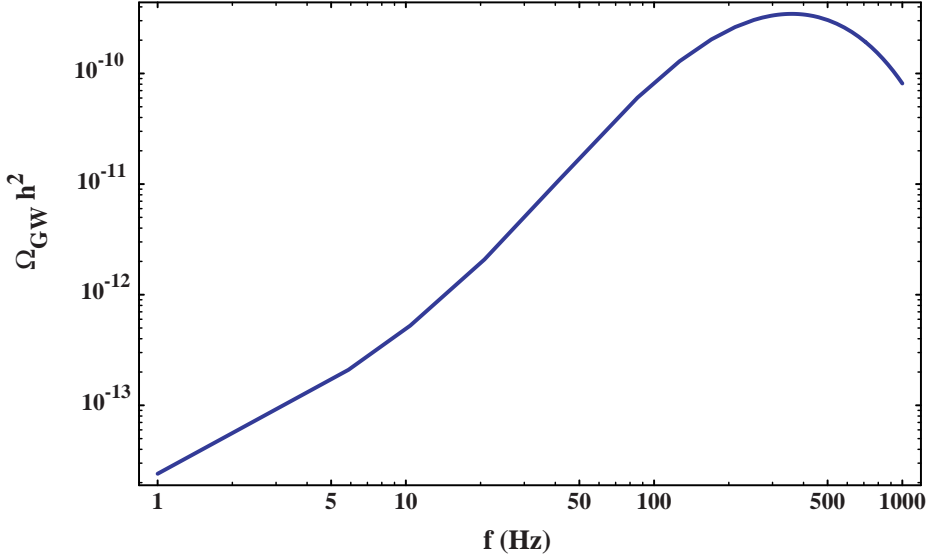}
\caption{Gravitational Wave background from core collapse supernovae in Model 0.
\label{gw710}}
\end{figure}

\begin{figure}[ht]
\centering
\includegraphics[width=0.75\textwidth]{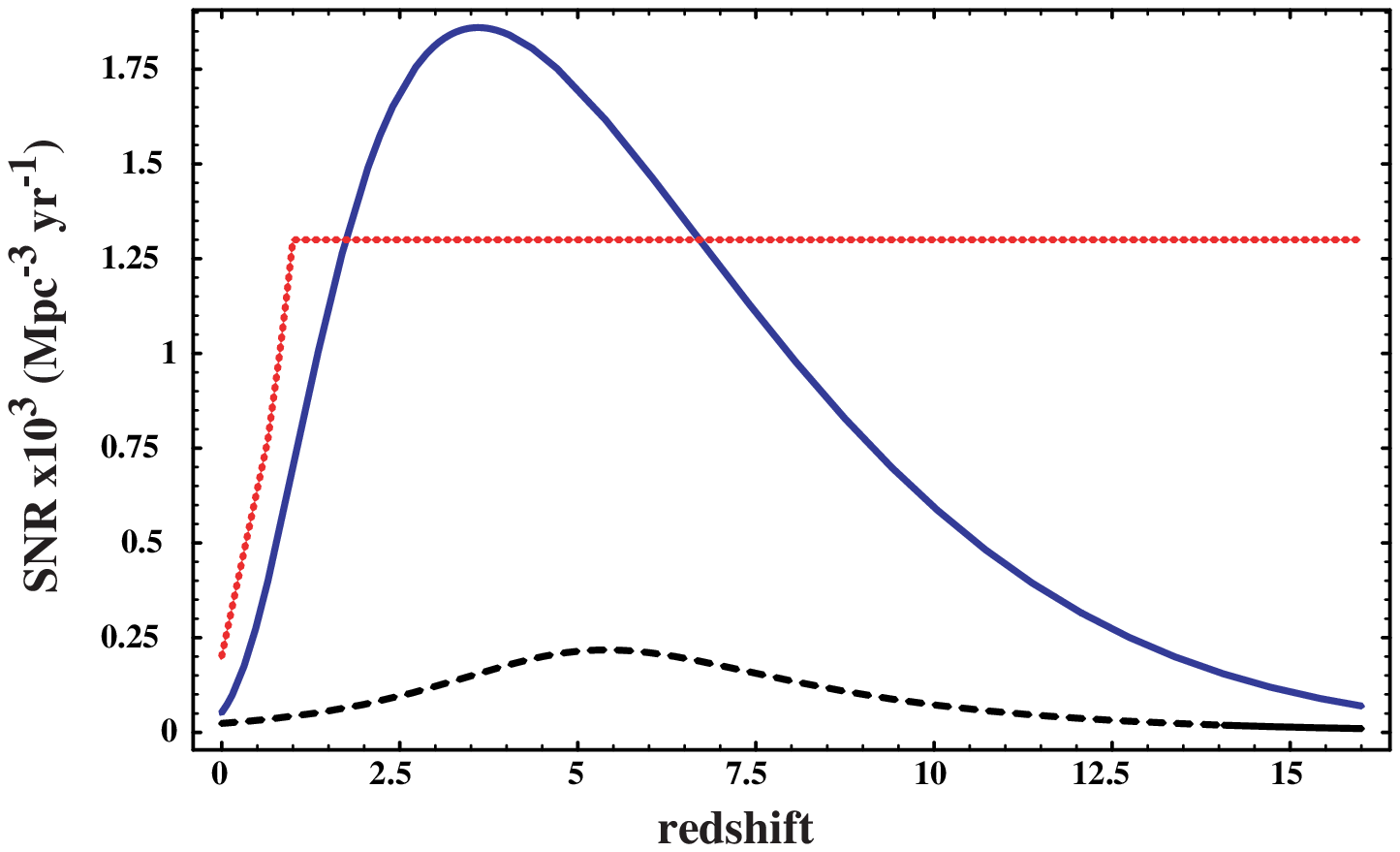}
\caption{Rates of core collapse supernovae considered in the calculation of the gravitational wave background for Model 0 (solid blue), from Buonanno {\it et al.} (dotted red)  \cite{bsrjm}, and using the Springel and Hernquist SFR from Araujo {\it et al.} (dashed black) \cite{araujoSH}.
\label{fig:SNR}}
\end{figure}

The differential energy density in gravitational waves from core collapse supernovae in Model 0 is shown in Fig.~\ref{gw710}. These stars collapse to form either neutron stars or black holes
with a mass equal to the mass of the progenitor's helium core. The spectrum peaks at a frequency of about 360 Hz, with $\Omega_{GW}h^2 = 3.5 \times 10^{-10}$. This spectrum should be similar to the upper curve in Fig.~4 of Ref.~\cite{bsrjm}, which peaks at roughly 300 Hz but with a maximum value about two orders of magnitude lower than that in Fig.~\ref{gw710}. One can see from Fig.~\ref{fig:SNR} that our supernova rate lies below theirs at high redshift and they are comparable at later times. A large supernova rate at high redshift serves to broaden the gravitational wave spectrum, but in this case it is not the cause of the disparity in peak heights, which is attributable primarily to differences in the IMFs. Their spectra were calculated under the assumption that all progenitors collapse to form neutron stars, whereas our Model 0 also includes stars that collapse to black holes. As can be seen from Eq.~\ref{IMF}, the number of stars collapsing to black holes in Model 0 is suppressed by an IMF that favors low masses, but stars emit $\sim 3$ orders of magnitude more energy in gravitational waves when the collapse yields a black hole than when it yields a neutron star due to the larger efficiency and remnant mass. As a result, the gravitational wave background for the normal mode of star formation is enhanced by two orders of magnitude over that in Buonanno {\it et al.} \cite{bsrjm}.

The background of gravitational waves from stars that collapse to black holes was calculated by Araujo {\it et al.}~\cite{araujoSH} assuming a Springel \& Hernquist \cite{SH} model of star formation. They find that for a Salpeter IMF defined for stars up to 125 M$_{\odot}$ and all stars with $25 < m <125$ collapsing to black holes with $M_r=\frac{1}{2}m$, the spectrum peaks at $\Omega_{GW}h^2 \approx 5 \times 10^{-9}$ at $f \approx 200$ Hz. This amounts to an expected background about an order of magnitude larger than that from Model 0. The Springel \& Hernquist SNR is actually smaller than any normal mode SNR considered here by a factor of $\sim8$ as one can see in Fig.~\ref{fig:SNR}, but they assume a much larger efficiency found in Ref.~\cite{sp85} of $\epsilon = 7 \times 10^{-4}$ and this remnant mass fraction is actually larger than any found in our Model 0.

\begin{figure}[ht]
\centering
\includegraphics[width=0.75\textwidth]{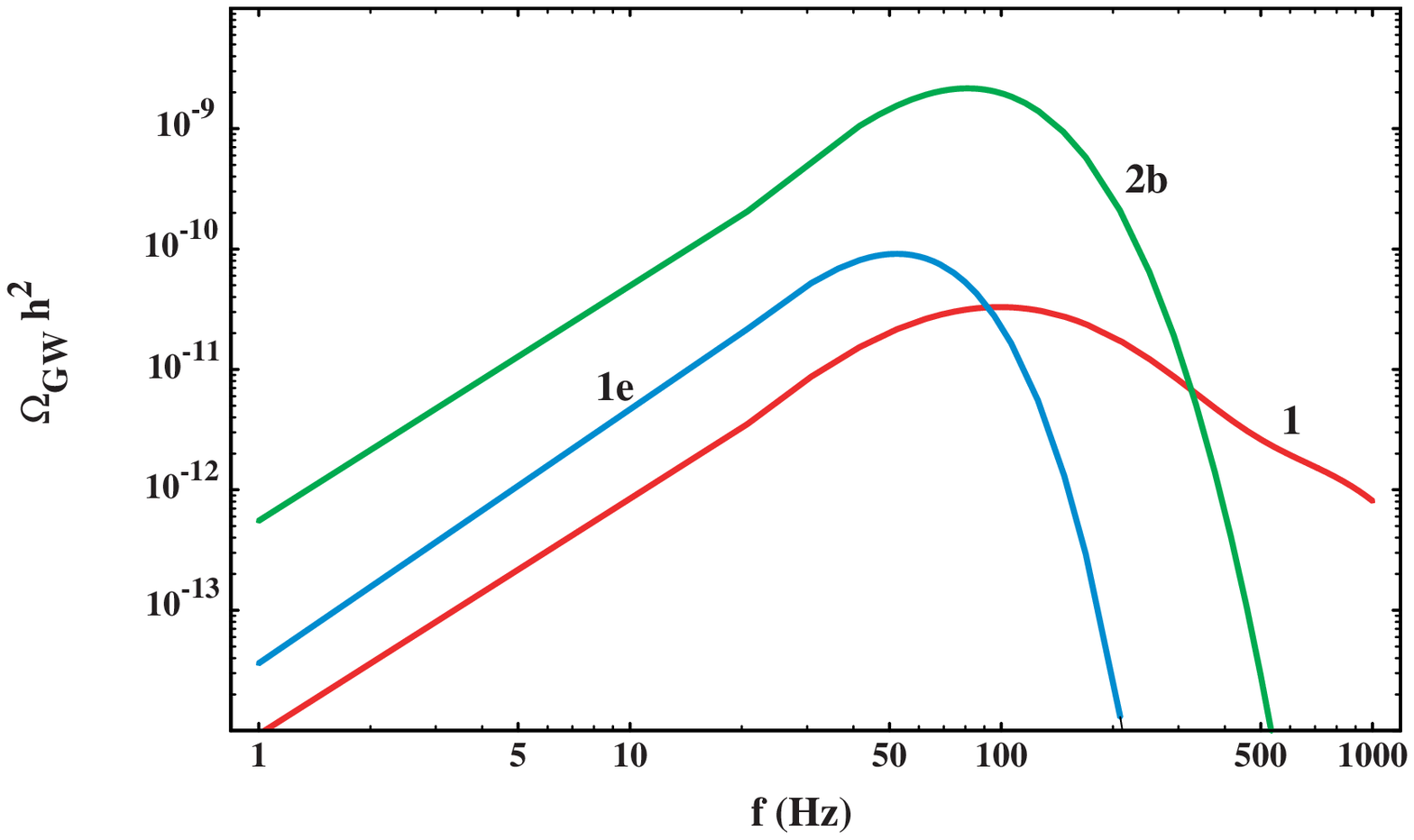}
\caption{Gravitational wave background from Population III core collapse supernovae in Models 1, 1e, and 2b.
\label{gwmassive}}
\end{figure}

Figure~\ref{gwmassive} shows the gravitational wave background from the massive
modes of Models 1, 1e, and 2b. In each case, the SFR has a different shape and amplitude. 
One can see that the spectrum from the massive mode of Model 2b is 
quite similar to the sharply peaked SFR case of Model 1e, with the 
peak location shifted due to star formation reaching maximum at different redshifts. 
As a burst, star formation in Model 1e occurs in a very limited redshift range around $z = 16$.
As a consequence, the frequency of gravitational waves does not extend much past 100 Hz.
In contrast, star formation in Model 2b does extend down to lower redshifts and hence
we expect the spectrum to extend to higher frequencies. 
Similarly, the gravitational wave spectrum from the massive mode of Model 1 extends to still higher frequencies due to significant star formation at redshifts between 3 and 5. The fact that maximum values of the energy density parameter span two orders of magnitude is attributable primarily to the differing SFRs seen in Fig.~\ref{fig:extremModel1SFRFracBZ}, but also to the differing remnant masses in Model 1 and 2b. Stars in Model 1 have much smaller initial masses and collapse to form black holes with a  mass equal to the progenitor's helium core, whereas stars in Model 2b, which are larger to begin with, collapse entirely to black holes. This entails larger bulk motions of matter, with $\Omega_{GW} \propto M_r$, as can be seen from Eqs.~\ref{omega}-\ref{energy}.

The spectrum from supernovae in Model 2b, which includes 300 M$_{\odot}$ progenitors, could be compared with the curves in Figure 8 of Ref.~\cite{bsrjm}. 
Although the frequencies at which the peaks occur are similar, the energy density obtained in our calculation is larger. The differing peak heights are attributed predominantly to the difference in the assumed supernova rates for the Pop III modes.
In Model 2b,  the baryon fraction in Pop III stars is $f_{III} = 7.0 \times 10^{-2}$. This is much larger than
the maximum value of $f_{III} =  10^{-3}$ assumed in \cite{bsrjm}, yielding $\Omega_{GW}h^2$ approaching $10^{-11}$. The supernova rate, and therefore the energy density in gravitational waves, scales linearly with $f_{III}$. In addition, we assume total collapse of 
massive stars leaving a remnant mass equal to the progenitor mass whereas Buonanno {\it et al.}~assume that 300 M$_{\odot}$ stars in collapse to form 100 M$_{\odot}$ black holes, and 
 $\Omega_{GW}h^2$ is linear in the remnant mass. Finally, Model 2b, is not a delta function 
 of 300 M$_\odot$ stars, but rather a power law distribution containing stars up to 500 M$_\odot$. 
These effects combine to enhance our peak height by a factor of about 300.
It is more difficult to compare Models 1 and 1e, as the IMF contains stars with masses less than 100
M$_\odot$.  Our fraction of baryonic matter in Population III stars is different for each model; $3.3 \times 10^{-3}$ for Model 1 and  $1.2 \times 10^{-2}$ for Model 1e.

\begin{figure}[ht!]
\centering
\includegraphics[width=0.75\textwidth]{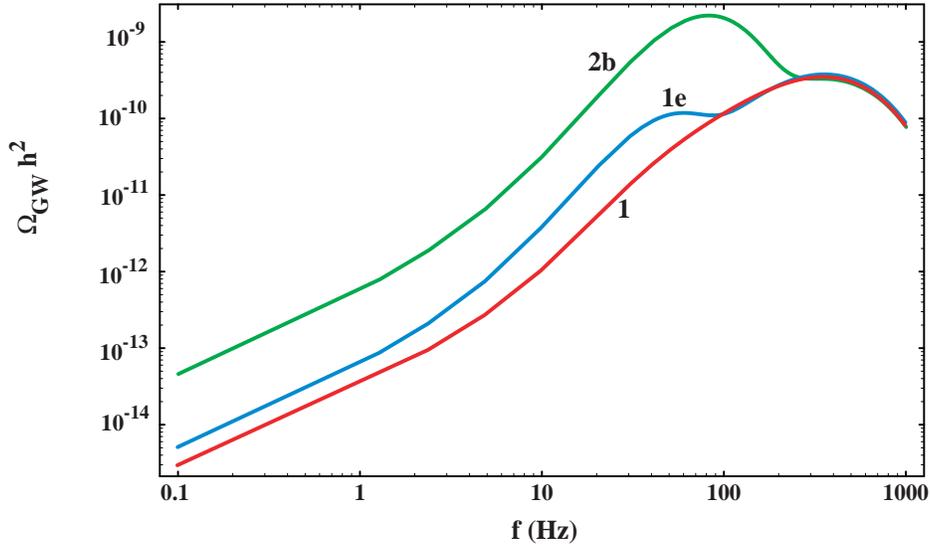}
\caption{Gravitational wave background from core collapse supernovae in Models 1, 1e, and 2b for both the normal and massive mode combined.
\label{gwtotal}}
\end{figure}

The total gravitational wave background for Models 1, 1e, and 2b, including the normal mode, is shown in Figure~\ref{gwtotal}. In the spectrum for Models 1e and 2b, one can clearly see the peak from the normal mode near 360 Hz and that from the massive mode at lower frequency. The massive mode contribution to the total background spectrum in Model 1 is not large enough to be seen here. Consequently, the spectra for Model 0 alone and Model 1 (including the normal mode) are identical. However, as discussed in Section~\ref{sec:SF models}, Model 0 alone could not provide a sufficient flux of ionizing photons for reionization at high redshift.

\begin{figure}[ht]
\centering
\includegraphics[width=0.75\textwidth]{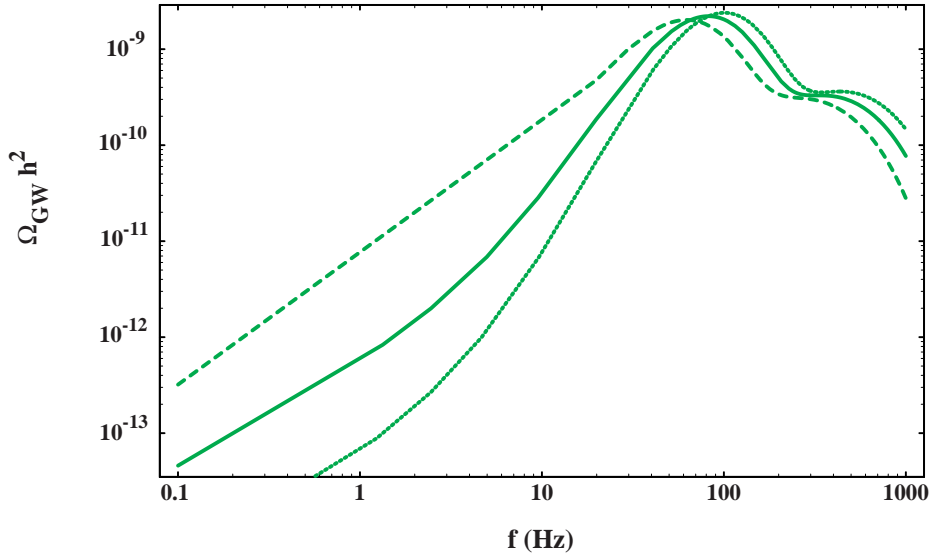}
\caption{Total gravitational wave background from core collapse supernovae in Model 2b including both the normal and massive modes. The three curves correspond to different values for the parameters $a$ and $b$. Our nominal case of $a = 200$ Hz and $b = 300$ Hz is shown as the solid curve.
The dashed (dotted) curve corresponds to  $a=b=250$ Hz ($a=150$ Hz and $b=350$ Hz).
\label{aandb}}
\end{figure}

As was mentioned in Section~\ref{sec:calc}, the shape of the spectrum of gravitational waves depends on the parameters $a$ and $b$ in Equation~\ref{efficiency}. While any variation in these parameters can have a dramatic effect on the resultant spectrum when the anisotropy parameter and neutrino luminosity are specified, the effects are strongly suppressed by our normalization scheme in which we specify the efficiency of gravitational wave production, $\epsilon$, for stars that collapse to black holes. The effect of varying the parameters $a$ and $b$ in the spectrum for Model 2b is shown in Figure~\ref{aandb}. 

\section{Detection}

As the backgrounds calculated here are the collective result of the collapse of all supernovae in the models considered, it is important to know whether the background is continuous. The ratio of the duration of each burst to the time between successive bursts is called the duty cycle, given by 
\beq DC = \int_0^{z_i} dR_{SN} \overline{\Delta \tau} (1+z)
\label{dc1}
\eeq
where $dR_{SN}$ is the differential rate of core collapse supernovae as seen from Earth and $\overline{\Delta \tau}$ is the average time duration of a single burst of gravitational wave emission~\cite{sfcfm}. One can approximate the duty cycle by choosing a fixed value for $\overline{\Delta\tau}$ or by  noting that the time duration of the burst corresponds to a frequency at emission, so $\overline{\Delta \tau} (1+z) \approx 1/f$, where $f$ is the observed frequency. Then the duty cycle can also be approximated by
\beq
DC \approx \frac{1}{f} \int_0^{z_i} dR_{SN} = \frac{1}{f} \int_0^{z_i} \int_{M_{min}}^{M_{max}} \phi(m) \frac{\psi(t-\tau(m))}{1+z} \frac{dV}{dz} dm dz,
\label{dc2}
\eeq
where $dV/dz = 4 \pi r^2(z) / H$ is the comoving volume element and $dr = (1+z) dt$~\cite{bsrjm}. The duty cycles approximated using these two methods are given in Table \ref{DC}. In column 2, the duration of the burst was assumed to be roughly the same for all supernovae, $\overline{\Delta \tau} \sim 1$ ms. As $DC < 1$, the background is not expected to be continuous. Using the method described by Eq.~\ref{dc2}, we find that the background is expected to be continuous for low frequencies ($f \lesssim 10$ Hz.), but that in each case individual bursts will be distinct (shot noise) throughout the peak of the spectrum.

\begin{table}[ht]
\begin{center}
\begin{tabular}{|c|c|c|}
\hline
Model & DC($\overline{\Delta\tau}=1$ ms) & $\sim$DC $\times f$ \\
\hline
0 & 0.18 & 37 \\
1 & 0.23 & 41 \\
1e & 0.47 & 54 \\
2b & 0.34 & 61 \\ 
\hline
\end{tabular}
\label{DC}
\caption{The duty cycles calculated for each of the models considered here. All include both the normal and massive modes, except Model 0, which is the normal mode only.}
\end{center}
\end{table}

\begin{figure}[ht]
\centering
\includegraphics[width=0.75\textwidth]{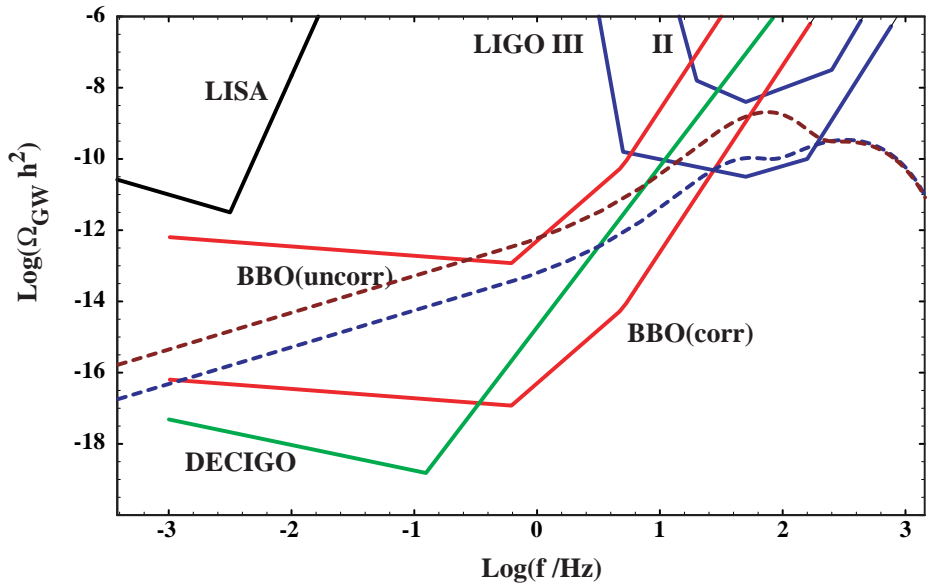}
\caption{Sensitivities of proposed and operating gravitational wave detectors BBO, DECIGO, LIGO II and III (correlated), and LISA expressed in terms of energy density \cite{tasi, seto}. The dashed lines are the expected gravitational wave backgrounds from Models 1e and 2b as in Figure~\ref{gwtotal}.
\label{explimits}}
\end{figure}

Finally we comment on the potential for current and future detectors to observe 
the gravitational signal from the first stars. Figure~\ref{explimits} shows the sensitivities of the LIGO II and III correlated ground based detectors and proposed sensitivities for LISA, the Big Bang Observer (BBO) and the Decihertz Interferometer Gravitational Wave Observatory (DECIGO). LISA, intended to search for gravitational waves from merging binaries as well as any primordial background, is designed to be sensitive to low frequencies $< 0.1$ Hz. Unfortunately, comparison of LISA's sensitivity curve with the dashed lines representing the expected backgrounds for Models 1e and 2b show that LISA will be incapable of detecting the backgrounds predicted here. DECIGO and, as a follow on to LISA, BBO have been proposed to search for the inflationary gravitational wave background at higher frequencies. Both antennas will have sufficient sensitivity to detect the gravitational wave background from cosmological supernovae at frequencies $f \lesssim 10$ Hz. Based on CMB temperature fluctuations, the upper limit to the gravitational wave background from inflation was estimated to be $\Omega_{GW}h^2 = 10^{-15}$~\cite{cooray}, which may be completely masked by that from stellar collapse down to frequencies of $10^{-1}$-$10^{-2}$ Hz. 

With a lead time of more than 20 years for BBO (and similarly for DECIGO), LIGO correlated detectors are the most immediate promising prospect for detection. Although LIGO II will not have the requisite sensitivity to detect the background from any model considered here, LIGO III correlated detectors will be able to probe the frequency range $f \sim 10^1$ - few $\times 10^2$ Hz.~to sufficiently small values of the energy density that the peak from the massive mode may be observed. When considering the frequency ranges covered by both LIGO III and BBO/DECIGO, it is possible that the entire spectrum below a few hundred Hz.~may be observed in the near future.

\section{Conclusions}

We have calculated the gravitational wave background from three different star formation histories that reproduce the observed chemical abundances and reionize the universe at high redshift. Each star formation history consists of a Population III mode of star formation coupled to a normal mode that reproduces the observed SFR for $z \lesssim 6$. The gravitational wave background was calculated assuming the amplitude given in Eq.~\ref{fh}. We obtained three distinct gravitational wave background spectra. In each case, there is a peak due to the normal mode near $f = 360$ Hz.~as well as a peak due to the massive mode at lower frequency. The location of the massive mode peak depends on the redshift range over which Population III star formation occurred; the earlier star formation reached maximum, the smaller the peak frequency.

The gravitational wave background from core collapse supernovae in our models is found to be large enough that it should be detected by the next generation of space-based laser interferometers, BBO and DECIGO, if not sooner by LIGO. The background will constitute a shot noise signal in the frequency range accessible with LIGO, however we expect the signal to be continuous for frequencies that will be probed by BBO and DECIGO. Within the next few decades, a detection of the gravitational wave background from cosmological supernovae will provide valuable information about the history of structure formation in the universe.

{\bf Acknowledgements}

We would like to thank G. Sigl for useful conversations.
The work of K.A.O., F.D. and E.V. was supported by the Project ``INSU - CNRS/USA", and the work of K.A.O. and P.S. was also supported in part by DOE grant
DE--FG02--94ER--40823.

\end{document}